\documentclass[12pt]{article}
\usepackage{amssymb}

\evensidemargin \oddsidemargin
\def\1{{\bf 1}}

\def\ot{\otimes}

\def\F{\mbox{$\cal F$}}
\def\bF{\mbox{$\overline{\cal F}$}}

\def \D {{\cal D}}
\def \H {{\cal H}}
\def \P {{\cal P}}
\def \W {{\cal W}}
\def \G {{\cal G}}
\def\O{\mbox{$\cal O$}}
\def\R{\mbox{$\cal R$\,}}
\def\S{\mbox{$\cal S$}}

\def \bV {\overline{V}_+}
\def \tW {\widetilde{W}}

\def\A{\mbox{$\cal A$}}
\def\hA{\mbox{$\widehat{\cal A}$}}

\newcommand{\tr}{\hat\triangleright}
\newcommand{\trc}{\triangleright}

\def\g{\mbox{\bf g\,}}

\def\b#1{{\mathbb #1}}
\def\nn{\nonumber \\}

\newcommand{\be}{\begin{equation}}
\newcommand{\ee}{\end{equation}}
\newcommand{\bea}{\begin{eqnarray}}
\newcommand{\eea}{\end{eqnarray}}
\newcommand{\ba}{\begin{array}}
\newcommand{\ea}{\end{array}}
\begin{document}

\title{On the consequences of twisted Poincar\'e symmetry upon
QFT on Moyal noncommutative spaces\footnote{Talk given at the
Symposium in honor of Wolfhart Zimmermann's 80th birthday, Ringberg
Castle, February 2008.}}

\author{Gaetano Fiore  \\      \and
        Dip. di Matematica e Applicazioni,
        V. Claudio 21, 80125 Napoli\\  and \\
        I.N.F.N., Sez. di Napoli, Complesso MSA, V. Cintia, 80126 Napoli
        }
\date{}

\maketitle

\abstract{We explore some general
consequences of a consistent formulation of relativistic quantum field theory (QFT) on the Gr\"onewold-Moyal-Weyl noncommutative versions of Minkowski space
with covariance under the {\it twisted Poincar\'e
group}  of Chaichian {\it et al} \cite{ChaKulNisTur04},
Wess \cite{Wes04}, Koch {\it et al} \cite{KocTso04}, Oeckl \cite{Oec00}.
We argue that a proper enforcement of the latter
requires  {\it braided} commutation relations between any pair of
coordinates $\hat x,\hat y$ generating two different copies of the
space, or equivalently a {\it $\star$-tensor product} $f(x)\star g(y)$ (in the
parlance of Aschieri {\it et al} \cite{AscBloDimMeySchWes05}) between any two functions depending on $x,y$. Then all differences
$(x\!-\!y)^\mu$ behave like their undeformed counterparts. Imposing
(minimally adapted) Wightman axioms
one finds that the $n$-point functions fulfill
the same general properties as on commutative space. Actually,
upon computation one finds
(at least for scalar fields) that the  $n$-point functions remain unchanged
as functions of the coordinates' differences both
if fields are free and if they interact (we treat interactions via time-ordered
perturbation theory). The main, surprising outcome seems a QFT physically equivalent to the undeformed counterpart (to confirm it or not one should
however first clarify the relation between $n$-point functions and observables, in particular $S$-matrix elements).

These results are mainly based on a joint work \cite{FioWes07} with J. Wess.
}

\vskip1cm \noindent - Preprint 32-2008 Dip. Matematica e
Applicazioni, Universit\`a di Napoli;\\
\noindent - DSF-17/08

\newpage

\section{Introduction}

The idea of spacetime noncommutativity is rather old. It goes back
to Heisenberg\footnote{Heisenberg proposed it in a letter to Peierls
\cite{Hei30} to solve the problem of divergent integrals in
relativistic quantum field theory. The idea propagated via Pauli to
Oppenheimer. In 1947 Snyder, a student of Oppenheimer, published the
first proposal of a quantum theory built on a noncommutative space
\cite{Sny47}.}. The simplest noncommutativity one can think of is
with coordinates $\hat x^\mu$ fulfilling the commutation relations
\be [\hat x^\mu,\hat x^\nu]=i\theta^{\mu\nu}, \label{cr} \ee where
$\theta^{\mu\nu}$ are the elements of a constant real antisymmetric
matrix. Relations (\ref{cr}) have appeared in the literature under
various names\footnote{Sometimes they are called {\it canonical},
since by applying a Darboux transformation to the coordinates
$\theta$ can be brought to canonical form (this depends only on
its rank). More often the names contain some combination of the
names of Weyl, Wigner, Gr\"onewold, Moyal. This is due to the
relation between canonical commutation relations and the
$\star$-product (or twisted product) of Weyl and Von Neumann, which
in turn was used by Wigner to introduce the Wigner transform;
Wigner's work led Moyal to define the socalled Moyal bracket
$[f\stackrel{\star}{,} g]\!=\! f \star g \!-\! g \star f$; the
$\star$-product in position space [in the form of the asymptotic
expansion of (\ref{starij}) with $x_i\!=\!x_j\!\equiv\! x$] first
appeared in a paper by Gr\"onewold.}. For brevity, we shall denote
these noncommutative spaces as Moyal spaces. For present purposes
$\mu=0,1,2,3$ and indices are raised or lowered through
multiplication by the standard Minkowski metric $\eta_{\mu\nu}$, so
as to obtain a deformation of Minkowski space. Clearly (\ref{cr})
are translation invariant, but not Lorentz-invariant (in 4
dimensions there is no isotropic antisymmetric 2-tensor
$\theta^{\mu\nu}$). We shall denote by $\hA$ the algebra``of
functions on Moyal space'', i.e. the algebra generated by $\1,\hat
x^\mu$ fulfilling (\ref{cr}). For $\theta^{\mu\nu}=0$ one obtains
the algebra $\A$ generated by commuting $x^{\mu}$.

Contributions to the construction of QFT on these spaces start in
1994-95 \cite{DopFreRob95}. A broad attention has been devoted to the
program in the last decade, with a number of different approaches.
By no means are they equivalent! Roughly speaking
I would divide them into the following three groups.

\subsubsection*{Doplicher-Fredenhagen-Roberts (DFR) approach}

This is field quantization in (rigorous) operator formalism on
Moyal-Minkowski space, with usual Poincar\'e transformations.
The pioneering works are \cite{DopFreRob95}, the main developments can be found
in \cite{BahDopFrePia02-05}. Relations (\ref{cr}) are
motivated by the interplay\footnote{The arguments elaborate the
well-known heuristic ones going back (as far as I know) to
Wheeler \cite{Wee65}.} of Quantum Mechanics and General Relativity in what
Doplicher calls the {\bf Principle of gravitational stability against
localization of events:}

\smallskip
{\it The gravitational field generated by the concentration of
energy required by the Heisenberg Uncertainty Principle to localise
an event in spacetime should not be so strong to hide\footnote{By
black hole formation.} the event itself to any distant observer -
distant compared to the Planck scale} \cite{Dop06,DopFreRob95}.

In the first, simplest version $\theta^{\mu\nu}$ are not fixed
constants, but central operators (obeying additional conditions)
which on each irreducible representation  become fixed constants
$\sigma^{\mu\nu}$, the joint spectrum of $\theta^{\mu\nu}$.
This allows to recover Lorentz covariance for the commutation relations.
However, it seems that when developing the interacting theory
the wished Lorentz covariance is sooner or later lost.
In more recent versions $\theta^{\mu\nu}$ is no more central, but
commutation relations remain of Lie-algebra type.

According to speculations heard in conference talks by Doplicher,
$\theta^{\mu\nu}$ could be finally related to the vacuum expectation value (v.e.v.) of $R^{\mu\nu}$,
which in turn should be influenced by the presence of matter quantum fields in
spacetime (through quantum equations of motions).

\medskip

Finally, we would like to mention the work \cite{GroLec07},
which although not stricly in the DFR framework, also is based
on a continuos family of fields labelled by the whole spectrum of noncommutative parameters $\theta^{\mu\nu}$, but has some overlap
also with the following two approaches. A generalization of the procedure
\cite{GroLec07} has been proposed in the very recent work \cite{BucSum08}, see also Buchholz's contribution to the present volume.

\subsubsection*{Path-integral quantization approach}

This was initiated by Filk in \cite{Fil96} and has been adopted by most theoretical physicists, including many string-theorists, especially
after the work \cite{SeiWit99}. Useful reviews are in \cite{DouNek01,
Sza03}.
The string-theorists' main motivation is that such models
should describe the low-energy effective limit of string
theory in a constant background $B$-field. Lorentz covariance
[or $SO(4)$ covariance, after Wick-rotation]  is lost, but this is expected in
effective string theory because of the special direction selected by
the $B$-field; only covariance under a subgroup \cite{AlvVaz03} of $SL(2,\b{C})$,
the corresponding little group, is preserved.

The (Euclidean) classical field action used in the path-integral
is deformed replacing products of fields by $\star$-products,
whence modified Feynman rules for perturbative QFT are derived.

\medskip
New complications seem to appear, like non-unitarity after naive
Wick-rotation when $\theta^{0i}\neq 0$ \cite{GomMeh00},
violation of
causality \cite{SeiSusTou00,BozFisGroPitPutSchWul03},
mixing of UV and IR divergences \cite{MinRaaSei00}
and subsequent non-renormalizability, alleged change of statistics, etc.  Some
of these problems, like non-unitarity  or the very
occurrence  of divergences \cite{BahDopFrePia02-05},  may be
due simply to  naive (and unjustified) applications of commutative QFT rules
(path-integral methods, Feynman diagrams, analytic continuation, etc) and could disappear adopting the sounder field-operator approach.
As for UV-IR mixing, while
planar Feynman diagrams  remain as the undeformed (apart from a
phase factor), in particular have the same UV divergences,
nonplanar Feynman diagrams which were UV divergent become finite for
generic non-zero external momentum, but diverge as the latter go to
zero, even with massive fields: these are the IR divergences.
As a dramatic effect,
infinitely many counterterms are necessary, making
these theories non-renormalizable.

As a cure to the UV-IR mixing problem Grosse, Wulkenhaar
\cite{GroWul03} and collaborators
 add a $x$-dependent harmonic potential terms (e.g. $\sim\Omega^2
x^2\varphi\!\star\!\varphi$ for a scalar field)
to the Lagrangian (for a review see Grosse's contribution to the present
volume, and references therein). Then the theory becomes renormalizable;
actually $\Omega^2
x^2\varphi\star\varphi$ is the only other marginal/relevant operator
in the renormalization group flow.
However these terms spoil the translation invariance of the theory.

Moreover, up to now no notion of Wick rotation between such QFT on
Moyal-Euclidean space and QFT on Moyal-Minkowski noncommutative space
has been found (there might be none).

\subsubsection*{Twisted Poincar\'e covariant approaches}

These recover Poincar\'e covariance in a deformed version,
following the observation  \cite{ChaKulNisTur04,Wes04,KocTso04,Oec00} that
(\ref{cr}) are {\it twisted Poincar\'e group} covariant.
Field quantization is done either in a path-integral
(on the Euclidean) or in an operator  approach. The latter
is the framework adopted in the present contribution; this is mainly based
on the joint work \cite{FioWes07} with J. Wess, who unfortunately
has recently passed away.

{\bf How to implement twisted Poincar\'e covariance in QFT} has been
subject of debate and different proposals
\cite{ChaPreTur05,Tur06,BalManPinVai05,
BalPinQur05,BalGovManPinQurVai06,BuKimLeeVacYee06,Zah06,LizVaiVit06,Abe06},
two main issues being whether
one should: {\it a)} take the $\star$-product of fields at different spacetime points; {\it b)}  deform the canonical commutation relations (CCR)
of creation and annihilation operators $a,a^\dagger$ for free fields.

Our answers to questions {\it a)}, {\it b)} are affirmative and related to each other. The first arises from a proper analysis of twisted Poincar\'e
transformations (section \ref{sect2}). In section \ref{sect3} we adapt
Wightman axioms to the noncommutative setting replacing all products by
$\star$-products and analyze the consequences for Wightman and Green's
functions. Motivated by the construction of normalizable states
generated by the application to the vacuum of smeared fields (here we
show why test functions in the Schwarz space are fine for smearing
- a point we only briefly mentioned in \cite{FioWes07}), we choose a setting where $\star$-products
involve also the (Fock space) operator part of the fields; for free fields
(section \ref{sect4}) this corresponds to choosing the second of the
two options which were found admissible in \cite{FioWes07} (they both lead
to a $\star$-commutator of the fields equal to the undeformed counterpart).
In section \ref{sect4} we also briefly describe how the time-ordered
perturbative computation of Green functions of a scalar
$\varphi^{\star n}$-interacting theory gives the same results as the
undeformed theory (the Feynman rules being unchanged).
In section \ref{sect5} we comment on what we can learn from these results,
on which aspects still need investigation, and draw the conclusions.

\section{Twisting Poincar\'e  group and Minkowski spacetime}
\label{sect2}

As already noted (\ref{cr}) are translation invariant, but not
Lorentz-invariant. In \cite{ChaKulNisTur04,Wes04,KocTso04,Oec00} it
has been recognized that they are however covariant under a deformed
version of the Poincar\'e group, namely a triangular
noncocommutative Hopf $*$-algebra $H$ obtained from the Universal
Enveloping algebra (UEA) $U{\cal P}$ of the Poincar\'e Lie algebra
${\cal P}$ by {\it twisting} \cite{Dri83}\footnote{In section 4.4.1
of \cite{Oec00} this was formulated in terms of the dual Hopf
algebra}. This means that (up to isomorphisms) $H$  and $U{\cal P}$
(extended over the formal power series in $\theta^{\mu\nu}$) have

\begin{enumerate}
\item  the same $*$-algebra and counit $\varepsilon$ (i.e. trivial
representation);

\item coproducts $\Delta, \hat\Delta$ related by
\be\ba{l} \Delta(g)\equiv \sum_I g^I_{(1)}\ot
g^I_{(2)}\:\:\longrightarrow\:\:
\hat\Delta(g)=\F\Delta(g)\F^{-1}\equiv\sum_I g^I_{(\hat 1)}\ot
g^I_{(\hat 2)}\label{coproductn} \ea\ee for any $g\!\in\! H\equiv
U{\cal P}$. Fixed $\hat\Delta$, the socalled twist $\F\!\in\! H\ot
H$ is not uniquely determined, but what follows does not depend on
its choice. The simplest is \vskip-.5cm \be\ba{l} \qquad
\qquad\quad\F\equiv \sum_I\F^{(1)}_I\ot\F^{(2)}_I:=
\mbox{exp}\left(\frac i2\theta^{\mu\nu}P_{\mu}\ot P_{\nu}\right).
                                                   \label{twist}
\ea\ee $P_{\mu}$ denote the generators of translations, and in
(\ref{coproductn}), (\ref{twist}), we have used Sweedler notation;
the $\sum_I$ may be a series, e.g. $\sum_I\F_I^{(1)}\!\ot\!\F_I^{(2)}$ is the series arising from the power expansion
of the exponential;

\item antipodes $S,\hat S$ related by a similarity transformation;
this is  trivial for the above $\F$, so $\hat S= S$.

\end{enumerate}

\noindent For readers not familiar with Hopf algebras, we recall
that the coproduct is the abstract operation by which one constructs
the tensor product of any two representations. For the cocommutative
Hopf algebra $U\g$ ($\g$ being a generic Lie algebra)
$$
\Delta(\1)\!=\!\1\ot\1,\qquad \qquad
g\in\g\to\Delta(g)=(g\ot\1+\1\ot g)\equiv g_1+g_2
$$
and $\Delta$ is extended  as a $*$-algebra map \be \Delta:U\g\to
U\g\ot U\g,\quad \qquad \Delta(ab)=\Delta(a)\Delta(b), \qquad
\Delta(a^*)=[\Delta(a)]^{*\ot *}.        \label{deltaprop}
\ee The extension is unambiguous, as
$\Delta\big([g,g']\big)=\big[\Delta(g),\Delta(g')\big]$ if
$g,g'\in\g$.  Also $\hat\Delta$ fulfills (\ref{deltaprop}),
$\hat\Delta(\1)\!=\!\1\ot\1$, as well as compatibility with $\epsilon$ and $*$
(as $\F$ is unitary). Then $\hat\Delta$ can replace $\Delta$ in
constructing the tensor product of two representations of $U\g$. The
antipode is the abstract operation by which one constructs the
contragredient of any representation; it is uniquely determined by
the coproduct, if it exists. In the present case, it is determined
by $S(g)=-g$ if $g\in\g$, $S(\1)=\1$, $S(ab)=S(b)S(a)$. Altogether,
the structures $(U\P,\cdot,*,\Delta,\epsilon,S)$,
$(H,\cdot,*,\hat\Delta,\epsilon,S)$ are examples of Hopf
$*$-algebras (here we have explicitly indicated the algebra product
by $\cdot$, but for brevity everywhere we shorten $a\!\cdot\!
b=ab$).

For $U\P$ a straightforward computation  gives
$$
\hat\Delta (P_\mu)=P_\mu\!\ot\!\1\!+\!\1\!\ot\! P_\mu=
\Delta(P_\mu),\qquad\quad\hat\Delta
(M_\omega)=M_\omega\!\ot\!\1\!+\!\1\!\ot\! M_\omega+
P[\omega,\theta]\!\ot\! P\neq\Delta (M_\omega),
$$
where we have set $M_\omega\!:=\!\omega^{\mu\nu}M_{\mu\nu}$ and used
a row-by-column matrix product on the right. The left identity shows
that the Hopf $P$-subalgebra remains undeformed and equivalent to
the abelian translation group  $\b{R}^4$. Therefore, denoting by
$\trc,\tr$ the actions of $U{\cal P},H$ (on $\A$ $\trc$ amounts to
the action of the corresponding algebra of differential operators,
e.g.   $P_{\mu}$ can be identified with $i\partial_{\mu}:=
i\partial/\partial x^{\mu}$), they coincide on first degree
polynomials $a,b$ in $x^\nu,\hat x^\nu$, \be P_{\mu}\trc
x^{\rho}=i\delta^{\rho}_{\mu}=P_{\mu}\tr \hat x^{\rho}, \qquad\quad
M_\omega\trc x^{\rho}=2i(x\omega)^\rho, \qquad\quad M_\omega\tr \hat
x^{\rho}=2i(\hat x\omega)^\rho,
   \label{Px}
\ee but $\trc,\tr$  differ on higher degree polynomials in $x,\hat
x$, as they are extended by the rules at the lhs of \bea &&\ba{l}
g\trc\! (ab)\! =\!\sum_I\big( g_{(1)}\!\trc a\big) \!\big(
g_{(2)}\!\trc b\big)\ea\label{Leibniz}\\[8pt] &&\ba{l} g\tr (\hat a\hat b)
\!=\!\sum_I\!\big( g^I_{(\hat 1)}\tr \hat a\big) \!\big( g^I_{(\hat
2)}\tr \hat b\big)\qquad\Leftrightarrow\qquad g\trc \!(a\!\star\! b)
\!=\!\sum_I\!\big( g^I_{(\hat 1)}\trc a\big)\!\star\! \big(
g^I_{(\hat 2)}\trc b\big) \ea \label{Leibnizn} \eea resp. involving
the coproducts $\Delta(g),\hat \Delta(g)$ (these resp. reduce to the
usual or a {\it deformed} Leibniz rule if  $g=P_\mu,M_{\mu\nu}$).
Moreover, $(g\trc a)^*= (Sg)^*\trc a^*$ as usual.
Summarizing, the $H$-module unital $*$-algebra $\hA$ is  obtained by
twisting the $U{\cal P}$-module unital $*$-algebra $\A$.

\subsubsection*{Several spacetime variables. Formulation through $\star$-products.}

For $n\ge 1$ we denote the $n$-fold tensor product algebra of $\A$
by $\A^n$ and $x^{\mu}\!\ot\!\1\!\ot...$,
$\1\!\ot\!x^{\mu}\!\ot\!...$,... respectively by $x^{\mu}_1$,
$x^{\mu}_2$, ...  In other words, $\A^n$ is the algebra of functions
of $n$ sets of Minkowski coordinates $x^{\mu}_i$, $i=1,2,...,n$. The
proper noncommutative deformation of $\A^n$ is the noncommutative
unital $*$-algebra $\hA^n$ generated by real variables $\hat
x^{\mu}_i$  fulfilling the commutation relations at the lhs of \be
[\hat x^{\mu}_i,\hat x^{\nu}_j]=\1 i\theta^{\mu\nu}
\quad\qquad\Leftrightarrow\quad\qquad [ x^{\mu}_i\stackrel{\star},
x^{\nu}_j]=\1 i\theta^{\mu\nu}. \label{summary} \ee Note that the
commutators are not zero for $i\neq j$; some authors erroneously
impose (\ref{summary}) only for $i=j$. Relations (\ref{summary}) are
compatible with the Leibinz rule (\ref{Leibnizn})$_1$, so as to make
$\hA^n$ a $H$-module $*$-algebra, and are dictated by the {\it
braiding} (see e.g. \cite{Maj95}) associated to the quasitriangular
structure $\R=\F_{21}\F^{-1}$ of $H$; here
$\F_{21}=\sum_I\F^{(2)}_I\ot\F^{(1)}_I$.

As $H$ is even triangular (i.e. $\R\R_{21}=\1^{\ot 2}$), an
essentially equivalent formulation of these $H$-module algebras is
in terms of $\star$-products derived from $\F$. Denote by
$\A^n_\theta$ the algebra obtained by endowing the vector space
underlying $\A^n$ with a new product, the $\star$-product, related
to the product in $\A^n$ by \be\ba{l} a\star
b:=\sum_I(\bF^{(1)}_I\trc a)  (\bF^{(2)}_I\trc b), \ea
\label{starprod} \ee \vskip-2mm\noindent with $\bF\equiv\F^{-1}$.
This encodes both the usual $\star$-product within each copy of
$\A$, and the ``$\star-$tensor product'' between different copies
\cite{AscBloDimMeySchWes05,AscDimMeyWes06}. As a result one finds
the isomorphic $\star$-commutation relations at the rhs of
(\ref{summary}) [this follows from computing $x^{\mu}_i\!\star\!
x^{\nu}_j$, which e.g. for the specific choice (\ref{twist}) gives
$x^{\mu}_i x^{\nu}_j\!+\!i\theta^{\mu\nu}/2$] and that
$\hA^n,\A^n_\theta$ are isomorphic $H$-module unital $*$-algebras,
in the sense of the equivalences (\ref{Leibnizn}), (\ref{summary}).

The $\star$-product (\ref{starprod}) can be extended from
polynomials $a,b$ to power series. More explicitly, on analytic
functions $a(x_i),b(x_j)$ (\ref{starprod}) reads
\be
a(x_i)\star
b(x_j)= \exp[\frac i2\partial_{x_i}\theta\partial_{x_j}]a(x_i)
b(x_j)                        \label{starij}
\ee
(for any 4-vectors $p,q$ we define $p\theta
q:=p_{\mu}\theta^{\mu\nu}q_{\nu}$), what must be
followed by the indentification $x_i\!=\!x_j$ {\it after} the action
of the bi-pseudodifferential operator $\exp[\frac
i2\partial_{x_i}\theta\partial_{x_j}]$ if $i\!=\!j$. Strictly
speaking, the last formula makes sense if $a,b$ belong to some
suitable subalgebra \cite{EstGraVar89}
 $\A^n{'}$ of the algebra of analytic functions
such that the $\theta$-power series is not only termwise
well-defined but also convergent. Clearly $\A^n{'}$ will not be
large enough for quantum-field-theoretic purposes. On the other
hand, if $a(x_i),b(x_j)\in\A^n{'}$ admit Fourier transforms $\hat
a(k_i),\hat b(k_j)$ then
\be
a(x_i)\!\star\! b(x_j)=\int\!\!\!d^4k
\!\int\!\!\!d^4q \,\hat a(k)\hat b(q) \exp[i(k\cdot x_i+q\cdot x_j-
k\theta q/2)]. \label{IntForm}
\ee
 This can be used as a {\it
definition} of a $\star$-product for $a(x),b(x)\!\in\!
L^1(\b{R}^4)\cap \widehat{L^1(\b{R}^4)}$, for
$a(x)\!\in\!\S(\b{R}^4)$ (Schwarz space) and $b(x)\in \S'(\b{R}^4)$
(the space of tempered distributions), or conversely, as well as for
$a(x),b(x)\in \S'(\b{R}^4)$ provided $i\!\neq\!j$. These are in fact
enough to reproduce all the product operations used in ordinary QFT,
with results reducing to the commutative ones for
$\theta^{\mu\nu}\!=\!0$.

Actually, for $i\!=\!j$ and some $a(x),b(x)\in \S'(\b{R}^4)$ it may
even happen that (\ref{IntForm}) is ill-defined for
$\theta^{\mu\nu}\!=\!0$, but well-defined \cite{GraVar88} (and thus
``regularized'') for $\theta^{\mu\nu}\!\neq\! 0$\footnote{For
instance, for $a(x)=\delta^4(x)=b(x)$ and invertible $\theta$ one
easily finds $a(x_i)\!\star\! b(x_j)=(\pi^4\det
\theta)^{-1}\exp[2ix_j\theta^{-1}x_i]$; in particular for $i\!=\!j$
the exponential becomes 1 by the antisymmetry of $\theta^{-1}$, and
one finds a diverging constant as $\det \theta\to 0$, cf.
\cite{GraVar88,EstGraVar89}. In \cite{GraVar88} the largest algebra
of distributions for which the $\star$-product is well-defined and
associative is determined. In \cite{EstGraVar89} the subalgebra of
analytic functions for which (\ref{starij}) gives an asymptotic
expansion of (\ref{IntForm}) is determined.}.

$\S(\b{R}^4)$ is a $*$-module of the $*$-algebra underlying both
$U\P,H$. As usual, the irreducible submodules are the eigenspaces of
the Casimir $p\!\cdot\! p$; one can endow those characterized by a
positive eigenvalue $m^2$ and a positive spectrum for $P^0$ by the
usual pre-Hilbert space structure. By completion, one obtains
unitary irreducible representations (irreps) of the $*$-algebra
underlying both $U\P,H$, that describe scalar particles.
(Generalized) eigenfunctions of $P_\mu$ or $M_{\mu\nu}$ exist
instead within  $\S'(\b{R}^4)$, which is a larger $*$-module  of the
$*$-algebra underlying both $U\P,H$. Unitary irreps describing
higher spin particles can be obtained in the standard way as some
$\b{C}^k\!\ot_{\b{C}}\!\S(\b{R}^4)$ or projective modules thereof
(spinor bundles, 4-vector bundles, etc). Summarizing, one obtains
the same \cite{ChaKulNisTur04} classification ({\it \`a la} Wigner)
of elementary particles as unitary irreps of either $U\P$ or $H$.

The generalization of the definition (\ref{IntForm}) to
functions/distributions depending nontrivially on several (possibly
all the) $x_i$ is straightforward. In particular the $\star$-product
$a\star b$ is well-defined for any $a\!\in\!\S(\b{R}^{4n})$ and
$b\in \S'(\b{R}^{4n})$ (or viceversa). Also $\S(\b{R}^{4n})$,
$\S'(\b{R}^{4n})$ are $*$-modules of the $*$-algebra underlying both
$U\P,H$. In fact, we shall need to embed them in an even larger
module $*$-algebra $\Phi^e$ of operator-valued
(instead of $c$-number valued) distributions.  The
action $\trc$ fulfills the ordinary (resp. deformed) Leibniz rule
(\ref{Leibniz}) [resp. (\ref{Leibnizn})$_2$] if $a,b$ are multiplied
(resp. $\star$-multiplied). This implies that the action of $U\P,H$
on tensor products of modules is constructed using the ordinary
(resp. deformed) coproduct.

\medskip
In the sequel we shall formulate the noncommutative spacetime only
in terms of $\star$-products and construct QFT on it replacing all
products by $\star$-products.

\bigskip\noindent
The {\bf differential calculus is not deformed}, as
$P_\mu\trc\partial_{ x^{\nu}_i}=0$ implies $\partial_{ x^{\nu}_i}
\star=\partial_{ x^{\nu}_i}=\star\partial_{ x^{\nu}_i}$:
$$
\partial_{
x^{\mu}_i}\star x^{\nu}_j=\delta^\nu_\mu\delta^i_j+ x^{\nu}_j\star
\partial_{x^{\mu}_i}
 \qquad\qquad
\left[\partial_{x^{\mu}_i}\stackrel{\star},\partial_{x^{\nu}_j}\right]=0
$$
($\hat\partial_{x^{\mu}_i}$ on $\hA^n$ is isomorphic). In the sequel
we shall drop the symbol $\star$ beside a derivative, as it has no
effect.
Also {\bf integration over the space is not deformed}:
\be
\int\!
d^4x \: a(x)\star b(x)=\int\! d^4x \: a(x)b (x) \label{int1}
\ee
[this holds in particular for all $a(x)\!\in\!\S(\b{R}^4)$ and
$b(x)\in \S'(\b{R}^4)$]. Stoke's theorem still applies. Using
(\ref{IntForm}) it is easy to check the property
\be
\int\! dx_i^4\,
b \star a(x_i)= b \star \int\! dx_i^4 \, a(x_i),\qquad\qquad\mbox{if
}b \mbox{ is independent of } x_i,               \label{starIntCR}
\ee analogous to the commutative conterpart [of course, if $a(x_i)$ is a
$c$-number valued function/distribution depending only on $x_i$, the
integral at the rhs is a $c$-number and the $\star$-product at the rhs
can be dropped]. Therefore, for our purposes we can consider
integration over any set of coordinates $x$ as an operation commuting
with the $\star$-product.

\bigskip
Let $a_i\!\in\!\b{R}$ with $\sum_ia_i=1$. An {\bf alternative set of real
generators of $\A^n_\theta$} is:
\be
\ba{l} \xi^{\mu}_i\!:=\! x^{\mu}_i\!-\! x^{\mu}_{i\!+\!1},\quad
i\!=\!1,...,n\!-\!1,\qquad\quad
 X^{\mu}\!:=\!\sum_{i=1}^na_i x^{\mu}_i. \quad
\ea                     \label{altgen}
\ee
All $\xi^{\mu}_i$ are translation invariant, $X^{\mu}$ is
not. It is immediate to check that
 $[X^{\mu}\!\stackrel{\star},\! X^{\nu}]=\1 i\theta^{\mu\nu}$, so $ X^{\mu}$
generate a copy $\A_{\theta,X}$ of $\A_{\theta}$, whereas $\forall
b\!\in\!\A_{\theta}^n$
\be \qquad \qquad\xi^{\mu}_i\star b=\xi^{\mu}_i b=b\star
\xi^{\mu}_i\qquad\Rightarrow\qquad
[\xi^{\mu}_i\stackrel{\star},b]=0, \label{startrivial} \ee
so $\xi^{\mu}_i$ generate a $\star$-central subalgebra
$\A_{\xi}^{n\!-\!1}$, and $\A^n_\theta\sim\!\A_{\xi}^{n\!-\!1}\!
\ot\A_{\theta,X}$. The $\star$-multiplication operators
$\xi^{\mu}_i\star$ have the same spectral decomposition on all
$\b{R}$ (including 0) as multiplication operators $\xi^{\mu}\cdot $
by classical coordinates; the joint eigenvalues make up a space-like, or a null, or a time-like $4$-vector, in the usual sense. Moreover,
$\A_{\xi}^{n\!-\!1},\A_{\theta,X}$ are actually $H$-module
subalgebras, with \be \ba{l} g\trc ( a\star b) \!=\!\sum_I\left(
g^I_{(1)}\trc a\right)\!\star\! \left( g^I_{(2)}\trc b\right) ,
\qquad\qquad a\!\in\! \A_{\xi}^{n\!-\!1},\quad
b\!\in\!\A_{\theta}^n,\quad g\!\in\! H,\ea\qquad
\label{UndefLeibniz} \ee i.e. {\it on $\A_{\xi}^{n\!-\!1}$ the
$H$-action is undeformed}, including the related part of the Leibniz
rule. [By (\ref{startrivial}) here $\star$ can be also dropped].

Inverting (\ref{altgen}), any set $x_i$ can be expressed as
a combination of the $n\!-\!1$ sets of $\star$-commutative variables $\xi_i$ and
the set $X$ of $\star$-noncommutative ones, e.g. if $X:= x_n$
then
$$
 x_i=\sum\limits_{j=i}^{n-1}\xi_j +  X.
$$
$X$ therefore behaves as parametrizing a ``global
noncommutative translation''.

\section{Revisiting Wightman axioms for QFT and their consequences}
\label{sect3}

As in Ref. \cite{Stro04} we divide the Wightman axioms
\cite{StrWig63} into a subset (labelled by {\bf QM})  encoding the
quantum mechanical interpretation of the theory, its symmetry under
space-time translations and stability, and a subset (labelled by
{\bf R}) encoding the relativistic properties. Since they provide
minimal, basic requirements for the field-operator framework to
quantization we try to  apply them to the above noncommutative space
(i.e. replacing everywhere products by $\star$-products) keeping the
QM conditions, twisting Poincar\'e-covariance R1 and
being ready to weaken locality R2 if necessary.

\bigskip \noindent
{\bf QM1.} 
The states are described by vectors of a (separable) Hilbert space
$\H$.

\medskip \noindent {\bf QM2.} 
The group of space-time translations $\b{R}^4$ is represented on
$\H$ by strongly continuous unitary operators $U(a)$: the fields
transform according to (\ref{transf}) with unit $A,U(A),
\Lambda(A)$. The spectrum of the generators $P_\mu$ is contained in
$\overline{V}_+ = \{p_\mu: p^2 \geq 0, \,p_0 \geq 0 \}$. There is a
unique  Poincar\'e invariant state $\Psi_0$, the {\em vacuum state}.

\medskip \noindent {\bf QM3.} 
The fields (in the Heisenberg representation) $\varphi^\alpha(x)$
[$\alpha$ enumerates field species and/or $SL(2, \b{C})$-tensor
components]  are operator (on $\H$) valued tempered distributions on
Minkowski space, with $\Psi_0$ a {\em cyclic} vector for the fields,
i.e. $\star$-polynomials of the smeared fields applied to $\Psi_0$
give a set $\D_0$ dense in $\H$.

For a single scalar field $\D_0$ is
spanned by
vectors of the form
of a finite sum \be \Psi_{\underline{f}} = f_0 \Psi_0 +
\varphi(f_1)\,\Psi_0 +
\varphi\left(f_2^{(1)}\right)\star\varphi\left(f_2^{(2)}\right)\Psi_0
+..., \,\,\,                               \label{D_0} \ee where
$f_j^{(h)}\! \in\! \S(\b{R}^4)$, $h\le j\le N<\infty$ and
$$
\varphi(f):=\int\!\! d^4 x\, f(x)\star
\varphi(x)\stackrel{(\ref{int1})}{=}\int\!\! d^4 x\, f(x)
\varphi(x).
$$
The (non-smeared) polynomials in the fields on
commutative space make up a subalgebra $\Phi$
of what we may call the {\it (extended) field algebra}
$\Phi^e=\left(\bigotimes_{i=1}^{\infty}\! \S'\right)\!\ot\!\O$,
where the first, second,...
tensor factor $\S'$ is understood as the space of distributions
depending on ${\bf x}_1,{\bf x}_2,...$ [the dependence on
${\bf x}_{h}$ of the polynomial appearing in (\ref{D_0})
being trivial for $h\!>\! N$], and $\O$ is the $*$-algebra of
linear operators on $\H$ (e.g. for free bosonic/fermionic fields $\O$ is a
Heisenberg/Clifford algebra with infinitely many modes). $\Phi^e$ also
is a $U\P$-module $*$-algebra. We should therefore $H$-covariantly $\star$-deform
the whole $\Phi^e$ into the corresponding $\Phi^e_\theta$ (see also
\cite{Fio08}).
In analogy with the  commutative case, we shall require
that within $\Phi^e_\theta$ fields $\star$-commute with $c$-number valued
functions/distributions $f$
\be
[ \, \varphi^\alpha(x)\stackrel{\star},
f(y)\,] \equiv  \varphi^\alpha(x)\star f(y) -f(y)\star
\varphi^\alpha(x)= 0.                      \label{fieldfunctcomrel}
\ee
For free (scalar) fields this was proposed in \cite{FioWes07} as
the second of two admissible options (we shall explicitly recall
how this works in section \ref{sect3}); this relation, together with
(\ref{starIntCR}), implies
\be\ba{l}
\Psi_{\underline{f}} \!=\! f_0 \Psi_0 \!+ \!\!\int\!\! d^4 x_1
f_1(x_1)\!\star\! \varphi(x_1)\Psi_0 +\!\!\int\!\! d^4 x_1
\!\!\int\!\! d^4 x_2 f_2(x_1,x_2)\!\star\! \varphi(x_1)\!\star\!
\varphi(x_2)\Psi_0 +\!..., \\[10pt]
f_j(x_1,...,x_j):= f_j^{(1)}\!(x_1)\star...\star
f_j^{(j)}\!(x_j),\ea\label{state}
\ee
so $\Psi_{\underline{f}}$ is characterized by the terminating
sequence $\underline{f}=(f_0, f_1, ...f_N)$. It is immediate to
check that the Fourier transform of $f_j$ differs from the commutative one
only by a phase factor,
$$
\tilde f_j(p_1,...,p_j)=\tilde f_j^{(1)}(p_1)...\tilde
f_j^{(j)}(p_j) \exp\left[\frac i2
\sum_{h=1}^{j}\sum_{k=h+1}^{j}p_h\theta p_k\right],
$$
and therefore $f_j\! \in\! \S(\b{R}^{4j})$. As on commutative space,
$\D_0$ is also dense in
the set $\D_1$ of all vectors of the form (\ref{state}) with $f_j\!
\in\! \S(\b{R}^{4j})$.

\bigskip
Taking v.e.v.'s we define the {\it Wightman
functions}
\be
\W^{\alpha_1,...,\alpha_n}(x_1,...,x_n):=
\left(\Psi_0,\varphi^{\alpha_1}(x_1)\star...\star\varphi^{\alpha_n}(x_n)
\Psi_0\right), \label{DefWig}
\ee
which are in fact distributions,
and (their combinations) the {\it Green's functions} \be
G^{\alpha_1,...,\alpha_n}(x_1,...,x_n)\!:=\!
\left(\Psi_0,T\!\left[\varphi^{\alpha_1}\!(
x_1)\!\star...\star\!\varphi^{\alpha_n}\!(x_n)\right]\!
\Psi_0\right) \ee where also {\it time-ordering } $T$ is defined as
on commutative space (even if $\theta^{0i}\neq 0$), e.g.
$$
T\!\left[\varphi^{\alpha_1}\!(x)\!\star\!\varphi^{\alpha_2}\!(y)\!\right]
\!=\!\varphi^{\alpha_1}\!(x)\!\star\varphi^{\alpha_2}\!(y)
\star\vartheta(x^0\!-\!y^0)\!
+\!\varphi^{\alpha_2}\!(y)\!\star\varphi^{\alpha_1}\!(x)
\star\vartheta(y^0\!-\!x^0)
$$
for $n\!=\!2$ ($\vartheta$ denotes the Heavyside function). This is
well-defined as  $\vartheta(x^0\!-\!y^0)$ is $\star$-central: the
$\star$-products preceding all $\vartheta$ could be dropped, by
(\ref{startrivial}).

\medskip
Arguing as for ordinary QFT (see \cite{StrWig63}) one finds that
QM1-3 (alone) imply exactly the same properties as on commutative
space:

\vspace{.3mm}\noindent  {\bf W1.} Wightman and Green's functions are
translation-invariant tempered distributions and therefore may {\it
depend only on the $\xi^{\mu}_i$}: \be\ba{rcl}
\W^{\alpha_1,...,\alpha_n}(x_1,...,x_n)&=&
W^{\alpha_1,...,\alpha_n}( \xi_1,..., \xi_{n\!-\!1}),\\[8pt]
\G^{\alpha_1,...,\alpha_n}(x_1,...,x_n) &=&
 G^{\alpha_1,...,\alpha_n}( \xi_1,..., \xi_{n\!-\!1}).
\ea\label{W1}\ee

\noindent{\bf W2.} ({\bf Spectral condition}) The support of the
Fourier transform $\tW$ of $W$ is contained in the product of
forward cones,   i.e. \be { \tW^{\{\alpha\}}(q_1, ...q_{n\!-\!1}) =
0,\qquad\mbox{if }\:\exists j:\quad q_j \notin \bV.} \ee

From (\ref{state}), (\ref{DefWig}) it follows that the scalar
product of vectors
$\Psi_{g_j}=\varphi\left(g_j^{(1)}\right)\star...\star\varphi\left(g_j^{(j)}\right)\Psi_0$,
$\Psi_{f_k}=\varphi\left(f_k^{(1)}\right)\star...\star\varphi\left(f_k^{(k)}\right)\Psi_0$
is given by
$$
(\Psi_{g_j},\Psi_{f_k})=\!\int\!\! d^{4j} x \!\!\int\!\! d^{4k} y\,
g_j^*(x_j,...,x_1)\star f_k(x_1,...,x_k)
\star\W(x_1,...,x_j,y_1,...,y_k)
$$
with $f_k,g_j$ defined as in (\ref{state}). Using (\ref{W1}) it is
straightforward to prove\footnote{The $\star$ between $\W$ and the rest is
ineffective by (\ref{startrivial})$_1$, (\ref{DefWig})$_1$.
Also the $\star$ between $g_j^*$ and $f_k$ is ineffective: going to the Fourier
transforms,  the
corresponding phase factor reduces to 1 when exploiting the presence
of the Dirac's $\delta$ in the equality $\tilde\W(p_1,...,p_n)=
(2\pi)^4\delta^4(\sum_i p_i)\tW( p_1,p_1\!+\!p_2,...,
p_1\!+\!...\!+\!p_n)$.} that in fact the previous
formula holds also without $\star$ (as on commutative space):
\be
(\Psi_{g_j},\Psi_{f_k})=\!\int\!\! d^{4j} x \!\!\int\!\! d^{4k} y\,
g_j^*(x_j,...,x_1) f_k(x_1,...,x_k)\W(x_1,...,x_j,y_1,...,y_k)
\label{intscalprod}
\ee
Using (\ref{intscalprod}) (and the analogous formulae for non-scalar fields)
we find

 \medskip\noindent{\bf W3.} $\W^{\{\alpha\}}$ fulfill the same {\bf
Hermiticity and Positivity} properties following from those of the
scalar product in $\H$ as in the theory on commutative space.

For instance, for the Wightman functions
of a single scalar field they reads as follows:
$[\overline{\W(x_1, ...,x_n)}]^*=\W(x_n, ...,x_1)$, and for all terminating sequences $\underline{f}=(f_0, f_1, ...f_N)$ with $f_j\! \in\! \S(\b{R}^{4j})$
\be
\big(\Psi_{\underline{f}},\Psi_{\underline{f}}\big)\equiv\sum_{j,k=1}^\infty\int
\!\! d^{4j} x \!\!\int \!\! d^{4k} y \,f^*_j(x_j, ...x_1) f_k(y_1,
...y_k)\,\W(x_1,...x_j, y_1, ...y_k)\, \geq 0.\ee

\bigskip
The {\it ordinary} relativistic conditions on QFT are:

\medskip\noindent{\bf R1.} ({\bf Lorentz Covariance}) $SL(2,
\b{C})$ is represented  on $\H$ by strongly continuous unitary
operators $U(A)$,   and under the Poincar\'e transformations $U(a,
\,A) = U(a)\,U(A)$ \be U(a,\! A) \,\varphi^\alpha(x)\,U(a,\!
A)^{-1}\! = S^\alpha_{\beta}(A^{-1}\!)\,\varphi^\beta\big(\Lambda(A)
x \!+\! a\big),\quad \label{transf} \ee with $S$ a
finite-dimensional representation of $SL(2, \b{C})$.

\medskip\noindent{\bf R2.} ({\bf Microcausality or locality}) The
fields either commute or anticommute at spacelike separated points
\be { [ \,
\varphi^\alpha(x),
\varphi^\beta(y)\,]_{\mp} = 0, \qquad\mbox{for}\,\,\,(x - y )^2 <
0.} \label{fieldcomrel0} \ee

\medskip
In {\it ordinary} QFT as a consequence of QM2,R1 one finds

\vspace{.5mm}\noindent {\bf W4.} ({\bf Lorentz Covariance of
Wightman functions})
 \be
 \W^{\alpha_1\!...\!\alpha_n}\! \big(\Lambda(A)x_1, ...,\Lambda(A)x_n\!\big)
\!=\!S^{\alpha_1}_{\beta_1}(A)\!...\! S^{\alpha_n}_{\beta_n}(A)
 \W^{\beta_1\!...\!\beta_n}(x_1, ...,x_n).\quad\label{LorCov}
\ee In particular, Wightman (and Green) functions of scalar fields
are  Lorentz invariant.

\medskip
R1 needs a ``twisted'' reformulation {\bf R1$_\star$}, which we
defer. Now, however R1$_\star$ will look like, it should imply that
$W^{\{\alpha\}}$ are $SL_\theta(2,\b{C})$ tensors (in particular
invariant if all involved fields are scalar). But, as the
$W^{\{\alpha\}}$ are to be built only in terms of $\xi^{\mu}_i$ and
other $SL(2,\b{C})$ tensors (like $\partial_{x^{\mu}_i}$,
$\eta_{\mu\nu},\gamma^\mu$, etc.), which are all annihilated by
$P_\mu\trc$,   $\F$ 
will act as the identity and $W^{\{\alpha\}}$ will transform under
$SL(2,\b{C})$  as for $\theta=0$. Therefore {\bf we shall require W4
also if $\theta\neq 0$} as a temporary substitute of R1$_\star$.

\bigskip
The simplest sensible way to formulate the $\star$-analog of
locality is

\vspace{.3mm}\noindent {\bf R2}$_\star$. ({\bf Microcausality or
locality}) The fields either $\star$-commute or $\star$-anticommute
at spacelike separated points
\be { [ \, \varphi^\alpha(x)\stackrel{\star},
\varphi^\beta(y)\,]_{\mp} = 0, \qquad\mbox{for}\,\,\,(x - y )^2 <
0.} \label{fieldcomrel}
\ee
This makes sense, as space-like
separation is sharply defined, and reduces to the usual locality
when $\theta=0$. Therefore we shall adopt it.
Whether there exist reasonable weakenings of R2$_\star$ is
 an open question also on commutative space, and the same
restrictions will apply.

Arguing as in \cite{StrWig63} one proves that  QM1-3, W4, R2$_\star$
are independent and compatible, as they are fulfilled by free fields
(see below): the noncommutativity of a Moyal-Minkowski space is
compatible with R2$_\star$! As consequences of R2$_\star$ one again
finds

\vspace{.5mm} \noindent{\bf W5.} ({\bf Locality}) if $(x_j -
x_{j+1})^2 < 0$ \be \W(x_1, ... x_j, x_{j+1}, ...x_n) =\pm \W(x_1,
...x_{j+1}, x_j, ...x_n). \ee
 \noindent{\bf W6}. ({\bf Cluster property}) For any
spacelike $a$ and for $\lambda \to \infty$ \be
 \W(x_1, ...x_j, x_{j+1} + \lambda a, ...,x_n + \lambda
a) \to \W(x_1, ...,x_j)\,\W(x_{j+1}, ...,x_n), \ee (convergence in
the distribution sense); this is true also with permuted $x_i$'s.

\medskip
Summarizing: our QFT framework is based on {\bf QM1-3, W4,
R2$_\star$} and the technical requirement (\ref{fieldfunctcomrel}), or alternatively on the constraints {\bf W1-6} for
$\W^{\{\alpha\}}$, exactly as in QFT on Minkowski space. We stress
that this applies for all $\theta^{\mu\nu}$, even if
$\theta^{0i}\!\neq\!0$, contrary to other approaches. Moreover, we have
just seen that (contrary to \cite{ChaMnaTurVer07}) we can keep the
Schwarz space $\S(\b{R}^4)$ as the space of
test functions for smearing the fields. We shall keep it as this
guarantees not only the separability of $\H$ but also that a finite
number of subtractions is enough to define field products at the same point,
i.e. essentially the possibility to renormalize the theory.
However we should note that, for given $f_j^{(h)}\in\S(\b{R}^4)$,
the states (\ref{D_0}) do not coincide with their undeformed
counterparts. We do not know whether this might have consequences
on observables (as $S$-matrix elements).

\section{Free or interacting scalar field}
\label{sect4}

As the differential calculus remains undeformed, so remain the
equation of motions of free fields. Sticking for simplicity to the
case of a scalar field of mass $m$, the solution of the Klein-Gordon
equation reads 
\be
\ba{l} \varphi_0(x)= \int \!d\mu(p)\, [e^{-ip\cdot x}\star a^p
+a_p^\dagger \star e^{ip\cdot x}\,] \ea \label{fielddeco}
\ee
where
$d\mu(p)=\delta(p^2\!-\!m^2)\vartheta(p^0)d^4p=
dp^0\delta(p^0\!-\!\omega_{\bf p})d^3{\bf p}/2\omega_{\bf p}$ is the
invariant measure ($\omega_{\bf p}\!:=\!\sqrt{{\bf p}^2+m^2}$).
Postulating the axioms of the preceding section, except  {\bf
R2$_\star$}, one can prove that up to a positive factor (which can
be always reabsorbed in a field redefinition) \be
 \ba{l}
 W(x\!-\!y)= 
 \int d\mu(p)e^{-ip\cdot (x\!-\!y)}
\\[8pt]
 G(x\!-\!y)=
  -i\int \frac{d^4p}{2\pi}\frac{e^{-ip\cdot
 (x\!-\!y)}}{p^2-m^2+i\epsilon},\ea \label{2W}\ee
and therefore coincides with the undeformed counterpart.
 Adding also {\bf R2$_\star$} one can prove the {\bf free
field commutation relation}
\be
\ba{l} [\varphi_0(x)\stackrel{\star},\varphi_0(y)]= 2\int d\mu(p)\:
\sin\left[p\!\cdot \!(x\!-\!y)\right]=:iF(x\!-\!y), \ea \label{freecomm}
\ee
{\bf coinciding with the undeformed one}. Applying
$\partial_{y^0}$ to (\ref{freecomm}) and setting $y^0\!=\!x^0$ [this
is compatible with (\ref{summary})] one finds {\bf the canonical
commutation relation} \be [\varphi_0(x^0,{\bf
x})\stackrel{\star},\dot\varphi_0(x^0,{\bf y})]= i\,\delta^3({\bf
x}-{\bf y}).                      \label{Cancomm} \ee As a
consequence of (\ref{freecomm}), {\bf the $n$-point Wightman
functions} not only fulfill W1-W6, but {\bf coincide with the undeformed ones}, i.e. vanish if $n$ is
odd  and are sum of  products of 2-point functions (factorization)
if $n$ is even. 

A $\varphi_0$ fulfilling (\ref{freecomm}) can be obtained assuming
$P_\mu\trc a^{\dagger}_p=p_\mu a^{\dagger}_p$, $P_\mu\trc a^p=-p_\mu
a^p$, so as to extend the $\star$-product law also to
$a^p,a^{\dagger}_p$, and plugging in (\ref{fielddeco})
 $a^p,a_p^\dagger$  satisfying
\be
\ba{l} a^{\dagger}_p
\!\star\! a^{\dagger}_q= e^{-i
 p\theta q}\,     a^{\dagger}_q \!\star\! a^{\dagger}_p, \qquad
 a^p\!\star\!  a^q= e^{-i p\theta q} \,    a^q \!\star\!  a^p,\\[8pt]
 a^p\!\star\!
a^{\dagger}_q=e^{i p\theta q} \,  a^{\dagger}_q \!\star\! a^p+
2\omega_{\bf p}\delta^3({\bf p}\!-\!{\bf q}),\\[8pt]
a^p\!\star\! e^{iq\cdot x} =e^{-ip\theta q} \,e^{iq\cdot x}\!\star\!
a^p, \qquad a^{\dagger}_p\!\star\! e^{iq\cdot x} = e^{ip\theta q}
\,e^{iq\cdot x}\!\star\! a^{\dagger}_p.       \ea\label{aa+cr'}
\ee
Note the nontrivial commutation relations between the $a^p,a^{\dagger}_p$
and $c$-number valued functions, but
$[\varphi_0(x)\stackrel{\star},f(y)]=0$ as in (\ref{fieldfunctcomrel}).
The first three relations
define an example of a general deformed Heisenberg algebra \cite{Fio95}
\be
\ba{l} a^q \star a^p= R^{qp}_{rs} \: a^s\star a^r,
\qquad\qquad a^{\dagger}_p \star a^{\dagger}_q= R^{sr}_{pq} \:
a^{\dagger}_r \star a^{\dagger}_s,  \\[8pt]
a^p \star a^{\dagger}_q=\delta^p_q+ R^{rp}_{qs} \: a^{\dagger}_r
\star a^s, \ea             \label{aa+cr"}
\ee
covariant under a
triangular Hopf algebra $H$. Here  $\R:=\F_{21}\F^{-1}$ is the
triangular structure of $H$, $\{|p\rangle\}$ is the generalized
basis of the 1-particle Hilbert space consisting of (on-shell)
eigenvectors of $P_\mu$, $\delta^p_q\!=\!2\omega_{\bf
p}\delta^3({\bf p}\!-\!{\bf q})$ is Dirac's delta (up to
normalization), $R^{pq}_{rs}:=\langle {\bf p}|\ot\langle {\bf q}|\R |
{\bf r}\rangle \ot | {\bf s}\rangle= e^{i p\theta
q}\delta^p_r\delta^q_s$.

Up to normalization of $R$, and with $p,q,r,s\in\{1,...,N\}$,
relations (25) are also identical to the ones defining the older
$q$-deformed Heisenberg algebras of \cite{PusWor89,WesZum90}, based
on a quasitriangular $\R$ in (only) the {\it $N$-dimensional}
representation of $H=U_qsu(N)$.

\medskip
{\bf Remark.} In \cite{FioWes07} we actually found also a
different (and maybe more intuitive) way to construct a free field
fulfilling (\ref{freecomm}). It amounts to: 1. introducing $a^p,a_p^\dagger$
satisfying
\bea
&& a^{\dagger}_p a^{\dagger}_q= e^{i
 p\theta'\! q}\,     a^{\dagger}_q  a^{\dagger}_p, \qquad
a^p  a^q\!=\! e^{i p\theta'\! q} \,    a^q  a^p,  \qquad a^p
a^{\dagger}_q\!=\! e^{-i  p\theta'\! q} \,  a^{\dagger}_q  a^p\!+\!2
\omega_{\bf p}\delta^3({\bf p}\!-\!{\bf q}), \nn [8pt]
&& \mbox{(with }\theta'\!=\theta\mbox{)},\qquad \qquad\mbox{and
}[a^p,f(x)]=[a^{\dagger}_p,f(x)]=0,     \label{aa+cr}
\eea
(so $c$-number valued functions/distributions keep commuting
with  $a^p,a_p^\dagger$), as adopted e.g. in
\cite{BalPinQur05,LizVaiVit06,Abe06}; 2. restricting
$\star$-multiplication only to the functions/distributions part
(i.e. elements of the extended $\A^n_\theta$) of the fields . Consequently, instead of (\ref{fielddeco}) the field decomposition reads
$\varphi_0(x)\!=\! \int\!\!d\mu(p)\, [e^{-ip\cdot x} a^p\!
+\!a_p^\dagger  e^{ip\cdot x}]$ with such $a^p,a_p^\dagger$.
This leads to the same properties {\bf W1-W6}.
However, as $\varphi(f)$ does no more depend on spacetime
coordinates $x$, the $\star$ in (\ref{D_0}) and (\ref{state})
becomes redundant, and we obtain
$\Psi_{\underline{f}} = f_0 \Psi_0 + \!\!\int\!\! d^4 x_1
f_1(x_1) \varphi(x_1)\,\Psi_0 +\!\!\int\!\! d^4 x_1
\!\!\int\!\! d^4 x_2 f_2(x_1,x_2) \varphi_0(x_1)
\varphi_0(x_2)\,\Psi_0 +...,$ with
$f_j(x_1,...,x_j) \!=\! f_j^{(1)}(x_1)... f_j^{(j)}(x_j)$.
As a result, scalar products  $(\Psi_{g_j},\Psi_{f_k})$
cannot be expressed in terms of Wightman functions as in
(\ref{intscalprod}), but in the form
\bea &&
(\Psi_{g_j},\Psi_{f_k})=\!\int\!\! d^{4j} x \!\!\int\!\! d^{4k} y\,
g_j^*(x_j,...,x_1) f_k(x_1,...,x_k)\W'(x_1,...,x_j,y_1,...,y_k) \nn
&& \W'(x_1,...,x_j,y_1,...,y_k):=
\left(\Psi_0,\varphi_0(x_1)...\varphi_0(x_j)\varphi_0(y_1)...\varphi_0(y_k)
\Psi_0\right)               \nonumber          
\eea
(with no $\star$-products in the definition of $\W'$, as in
 \cite{BalGovManPinQurVai06}]).
The distributions $\W'$ do not fulfill all the properties W1-W6
(except of course in the undeformed case $\theta'=0$).
We also briefly consider some consequences of choosing $\theta'\neq\theta$
in (\ref{aa+cr})
($\theta'=0$ gives CCR among the $a^p,a_p^\dagger$, assumed in most
of the literature, explicitly \cite{DopFreRob95} or implicitly, in
operator \cite{ChaPreTur05,ChaMnaNisTurVer06} or in path-integral
approach to quantization) together with
$\varphi_0(x)\!=\! \int\!\!d\mu(p)\, [e^{-ip\cdot x} a^p\!
+\!a_p^\dagger  e^{ip\cdot x}]$ and definition (\ref{DefWig})$_1$
for the Wightman functions. One finds the non-local
$\star$-commutation relation
$$
\varphi_0(x)\star\varphi_0(y)=e^{i
\partial_x(\theta-\theta')
\partial_y} \varphi_0(x)\star\varphi_0(y)+i\,F(x-y),
$$
and the corresponding (free field) Wightman functions violate W4,
W6, unless $\theta'=\theta$.

\bigskip
Going back to our framework,
we now define {\bf normal ordering} as a $\A_{\theta}^n$-bilinear map of
field algebra into itself such that $(\Psi_0,:\!M\!:\Psi_0)=0$ fo
any field polynomial $M$, in particular $:\!\1\!: \:=0$. Applying it
to (\ref{aa+cr}) we find that it is consistent to define
$$
:\!a^p\!\star\!
a^q\!: \:=\!a^p\!\star\! a^q, \quad :\!a^{\dagger}_p\!\star\! a^q\!:
\:=\!a^{\dagger}_p\!\star\! a^q, \quad :\!a^{\dagger}_p\!\star\! a^{\dagger}_q\!:
\:=\! a^{\dagger}_p\!\star\! a^{\dagger}_q, \quad :\! a^p\!\star\!
a^{\dagger}_q\!: \: \:=\! a^{\dagger}_q\!\star\!  a^p e^{-i p\theta q}
$$
(note the phase).
 More generally, by definition in any monomial this map
reorders all  $a^p$ to the right of
all $a_q^{\dagger}$ introducing a  $e^{-i q\theta p}$ for each flip
$a^p \leftrightarrow a_q^{\dagger}$.
For $\theta=0$ the map reduces to the undeformed normal ordering.

As a result, one finds that the v.e.v. of any normal-ordered
$\star$-polynomial of fields is zero, that normal-ordered $\star$-products
of fields can be obtained from $\star$-products by the undeformed pattern of
subtractions,
and that {\bf the same   Wick theorem} as in the undeformed case holds.
Applying {\bf time-ordered perturbation theory} to an interacting
field again one can heuristically derive \cite{FioWes07},
through the same arguments used on commutative space,
the Gell-Mann--Low formula
\be
 G(x_1,...,x_n)  =\frac{
\left(\Psi_0,T\left\{\varphi_0(x_1)\star ...\star
\varphi_0(x_n)\star \exp\left[ -i\lambda \int dy^0  \
H_I(y^0)\,\right]\right\} \Psi_0\right)} {\left(\Psi_0,T\exp\left[
-i \int dy^0 \ H_I(y^0)\,\right]\Psi_0\right)} \label{formal} \ee
(which is rigorously valid under the assumption of
asymptotic completeness, $\H=\H^{in}=\H^{out}$).
Here $\varphi_0,H_I(x^0)$ denote the free ``in'' field (i.e. the
incoming field) and the interaction Hamiltonian in the interaction
representation, e.g. \be H_I(x^0)= \lambda \int\! d^3 x\
:\varphi^{\star m}_0(x)\!: \,\star, \qquad\qquad \varphi^{\star
m}_0(x) \equiv\underbrace{\varphi_0(x) \star ...
\star\varphi_0(x)}_{m\mbox{ times}}.           \label{HI}
\ee
Thus \cite{FioWes07} one finds that the {\bf Green functions (\ref{formal})
coincide with the undeformed ones} (at least perturbatively). They
can be computed by Feynman diagrams with the undeformed Feynman
rules, and the theory can be regularized and renormalized in the standard ways.

\section{Conclusions. What do we learn?}
\label{sect5}

Although various approaches to relativistic
QFT on Moyal-Minkowski space have been proposed, there
is still no generally accepted one. Operator-based approaches
look safer starting points, but twisting or not the Poincar\'e group, and doing it properly, makes the results radically different.

We have claimed here that a sensible theory with twisted Poincar\'e seems possible and avoids all complications (IV-UR,
causality/unitarity violation, statistics violation, cluster
property violation, loss of spacetime symmetry,...).
It naturally involves a compensation of operator ($a,a^\dagger$)
and spacetime noncommutativities,
so that the free field $\star$-commutators coincide with the undeformed ones.

The surprising and probably
disappointing fact is that also the corresponding $n$-point functions, expressed as functions of the coordinates' differences, coincide with the undeformed ones.
The natural consequence seems that no new physics, nor a more satisfactory formulation of the old one (e.g. by an inthrinsic UV regularization) is obtained
(at least for scalar fields),
although this can be confirmed only upon clarifying the relation between $n$-point functions and observables, in particular $S$-matrix elements.

Nevertheless we think that we can learn quite much from trying to understand
the reasons of these surprising results, which are in
striking contrast with the ones found in most of the literature,
as well as from using our approach as a laboratory for:
\begin{enumerate}
\item  searching and testing equivalent formulations of QFT on NC spaces:
Wick rotation into EQFT, path integral quantization, etc.;

\item clarifying notions such as asymptotic
states, spin-statistics, CPT, etc., on  noncommutative spaces;

\item properly formulating covariance properties of fields under
twisted symmetries (R1$_\star$), and clarify their connection to the
ordinary ones;

\item properly formulating gauge field theory on noncommutative spaces.
\end{enumerate}

\subsection*{Acknowledgments}
I would like to thank Prof.'s W. Zimmermann, E. Seiler and K. Sibold
for the very kind invitation to the ``Zimmermannfest 08'' conference,
and for the warm atmosphere experienced there.

\end{document}